\documentstyle[prb,aps,epsf]{revtex}
\begin{document}
\draft \title {\Large\bf THERMALLY ACTIVATED HALL CREEP OF FLUX LINES
  FROM A COLUMNAR DEFECT} \author
{{\bf D. A. Gorokhov and G. Blatter}\\
  {\it Theoretische Physik, ETH-H\"onggerberg, CH-8093 Z\"urich,
    Switzerland}} \maketitle
\begin{abstract}
  
  We analyse the thermally activated depinning of an elastic string
  (line tension $\epsilon$) governed by Hall dynamics from a columnar
  defect modelled as a cylindrical potential well of depth $V_{0}$ for
  the case of a small external force $F.$ An effective 1D field
  Hamiltonian is derived in order to describe the 2D string motion.
  At high temperatures the decay rate is proportional to
  $F^{{5}/{2}}T^{-{1}/{2}} \exp{\left [{F_{0}}/{F}-{U(F)}/{T}\right
    ]},$ with $F_{0}$ a constant of order of the critical force and
  $U(F) \sim{\left ({\epsilon V_{0}}\right )}^{{1}/{2}}{V_{0}/{F}}$
  the activation energy.  The results are applied to vortices pinned
  by columnar defects in superclean superconductors.
\end{abstract}
\vskip-0cm\hskip1.5cm{preprint ETH-TH/97-20,
accepted for publication in Phys. Rev. B}
\vskip0.0cm\hskip1.5cm
e-mail: gorokhov@itp.phys.ethz.ch


\section{Introduction}
\label{Introduction}

The search for new mechanisms of pinning is both scientifically
challenging as well as an important problem to be studied in view of
technological applications of high-$T_{c}$ superconductors.  Even if
the transport current is less than the critical one, energy
dissipation takes place due to quantum or thermally activated creep of
vortices\cite{Blatter}.  Recent measurements of the critical current
and of the magnetization relaxation rate in layered high temperature
superconductors show that columnar defects produced by irradiation
with heavy ions can strongly suppress the vortex motion\cite{experim}.
In this paper we present a theoretical study of the thermally
activated depinning of a vortex from a columnar defect in the presence
of a small transport current (classical creep).

The inverse life-time $\Gamma$ of a metastable state can be written in
the form $\Gamma=A e^{-{S}/{\hbar}}$, where $S$ is the Euclidean
action along the extremal trajectory and $A$ is the prefactor
determined by the fluctuations around the saddle-point solution (see
Ref.~\cite{Haenggi} for a general review concerning the decay of
metastable states). At low temperatures the saddle-point solution is
time-dependent and, consequently, $S$ depends on the dynamics of the
system (quantum creep), whereas at high temperatures the calculation
of $S$ alone does not involve the dynamics (classical creep). Here, we
concentrate on high temperatures but go beyond the usual exponential
accuracy by calculating the prefactor $A$, a task which does involve
the dynamics of the flux lines as well.

In high-$T_{c}$ superconductors the dynamics of vortices may be
dominated by either the dissipative or Hall term in the equation of
motion.  Microscopic calculations of the dynamic
constants\cite{Kopnin} show that the ratio of Hall and dissipative
coefficients ${\alpha}/{\eta}$ is approximately equal to
${\omega}_{0}{\tau}$, with ${\omega}_{0}$ and $\tau$ the level spacing
between localized Caroli-de Gennes-Matricon states in the core and the
relaxation time, respectively.  Recent experimental studies on
90\thinspace K crystals of YBCO (see Ref.\cite{Matsuda}) have been
interpreted as providing evidence for the superclean limit, with
${\omega}_{0}\tau\sim 15$ below 15\thinspace K, in which case the
contribution of dissipative forces can be neglected in this regime.

Quantum depinning of flux lines governed by Hall dynamics from a
columnar defect has been considered by Sonin and Horovitz\cite{Sonin}
for the case of a small external force.  For pancake vortices this
problem has been studied by Bulaevskii, Larkin, Maley, and
Vinokur\cite{Bulaevskii}.  The case $j_{c}-j\ll j_{c}$ has been
investigated by Chudnovsky, Ferrera, and Vilenkin\cite{Chudnovsky} and
by Morais-Smith, Caldeira, and Blatter\cite{Morais-Smith}.  On the
other hand, the problem of the depinning of a {\it massive} string
from a linear object has been solved by Skvortsov in the whole
temperature range\cite{Skvortsov} and the thermal depinning of a flux
line governed by dissipative dynamics has been considered by Kr\"amer
and K\'ulic\cite{Kramer1}.

The present paper is organized as follows: In section \ref{1D} we
reduce the two-dimensional problem of the string motion to an
effective one-dimensional problem. In section \ref{dr} the decay rate
of a trapped string is calculated. In section \ref{con} the results
are applied to vortices in superclean superconductors.

\section{1D effective Hamiltonian}
\label{1D}

Let us consider a string which is pinned by a columnar defect in the
presence of an external force. Both, the cylindrical defect as well as
the vortex are directed along the $c$-axis of the anisotropic
superconductor and the external magnetic field is supposed to be
sufficiently small, such that the interaction between the vortices can
be neglected.  Furthermore, we consider the situation where each
vortex is pinned by an individual defect, i.e., the concentration of
defects is assumed to be larger than that of vortices.  The free
energy density of the string describing this situation is
\begin{equation}
G[{\bf u}]=
\frac{\epsilon}{2}{\left (\frac{\partial {\bf u}}
{\partial z}\right )}^{2} + U({\bf u}),
\label{1}
\end{equation}
where
\begin{equation}
U ({\bf u})=
V_{cyl}\left ({\sqrt{u_{x}^{2}+u_{y}^{2} }}
\enspace\right )+V_{ext}(u_{x}).
\label{pott}
\end{equation}
Here, $\epsilon$ is the elasticity of the string and $u_{x}$ and
$u_{y}$ describe its displacement along the $x$ and $y$ directions,
with the $z$-axis chosen parallel to the defect. In Eq.~(\ref{pott}),
$V_{cyl}\left (\sqrt{{u_{x}}^{2}+{u_{y}}^{2}}\right)$ denotes the
cylindrical pinning potential and $V_{ext}=-Fu_{x}$ is the forcing
potential.  The function $V_{cyl}\left (r\right )$ is supposed to be
monotonously increasing and restricted from below and above.  We
assume boundary conditions ${\bf u}\left({\pm L}/{2}\right)=0$.

If $F\not= 0,$ the state of the vortex becomes metastable. Our main
goal is to investigate the decay rate as a function of temperature $T$
and force $F,$ where the external force is assumed to be small.  The
Lagrangian of the vortex (in real time dynamics) can be written as
\begin{equation}
L[u_{x},u_{y}]=\int_{-{L}/{2}}^{+{L}/{2}}dz
\left (
{\alpha} u_{y}\frac{\partial u_{x}}{\partial t}-G[u_{x},u_{y}]
\right ),
\label{real}
\end{equation}
where $\alpha$ is the Hall coefficient.  The corresponding equations
of motion take the form
\begin{equation}
\alpha\frac{\partial u_{x}}{\partial t}=
\frac{\partial U}{\partial u_{y}}-
\epsilon\frac{{\partial}^{2}u_{y}}{\partial z^{2}},
\label{u1}
\end{equation}
\begin{equation}
\alpha\frac{\partial u_{y}}{\partial t}=
-\frac{\partial U}{\partial u_{x}}+
\epsilon\frac{{\partial}^{2}u_{x}}{\partial z^{2}}.
\label{u2}
\end{equation}
Eqs.~(\ref{u1}) and (\ref{u2}) can be formulated as the equations of
motion of the 1D Hamiltonian density
\begin{equation}
H = \left [U\left ({x},\frac{p}{\alpha}\right )+\frac{\epsilon}{2}
{\left (\frac{\partial {x}}{\partial z}\right )}^{2}
+\frac{\epsilon}{2{\alpha}^{2}}
{\left (\frac{\partial p}{\partial z}\right )}^{2} \right ],
\label{hamm}
\end{equation}
where we have used the definitions ${x}\equiv u_{x}$ and
$p\equiv\alpha u_{y}.$ Using the variational procedure for the
Hamiltonian density $H$ we obtain
\begin{equation}
{\dot x}=\frac{\partial H}{\partial p}-\frac{\partial }{\partial z}
\frac{\partial H}{\partial p_{z}}=
\frac{\partial U(x,{p}/{\alpha})}{\partial p}-
\frac{\epsilon}{{\alpha}^{2}}\frac{{\partial}^{2}p}{\partial z^{2}},
\label{q1}
\end{equation}
\begin{equation}
{\dot p}=-\frac{\partial H}{\partial x}+\frac{\partial}{\partial z}
\frac{\partial H}{\partial x_{z}}=
-\frac{\partial U(x,{p}/{\alpha})}{\partial x}+
{\epsilon}\frac{{\partial}^{2}x}{\partial z^{2}}.
\label{q2}
\end{equation}
and one can easily see that Eqs.~(\ref{q1}) and (\ref{q2}) are
equivalent to Eqs.~(\ref{u1}) and (\ref{u2}) with $x=u_{x}$ and
$p=\alpha u_{y}$.  Thus we have reduced the 2D problem of the motion
of a vortex governed by Hall dynamics with an action given by
(\ref{real}) to the 1D problem of the motion of a vortex described by
the effective action $\int dz dt \left ( p{\dot x}-H \right ),$ with
$H$ given by Eq.~(\ref{hamm}).  The above idea has been introduced by
Volovik\cite{Volovik} for the tunneling of 2D vortices in a liquid
helium film, the motion of which is equivalent to that of a massless
particle in a magnetic field. Later, this idea has been used by
Feigel'man, Geshkenbein, Larkin, and Levit\cite{Feigel'man} within the
context of vortex tunneling in superclean high-${T}_{c}$
superconductors.  Chudnovsky, Ferrera, and Vilenkin\cite{Chudnovsky}
have generalized this idea for the description of the depinning of a
flux line governed by Hall dynamics from a columnar defect near
criticality. In their case $U(u_{x},u_{y})=a u_{x}^{2}/{2}-b
u_{x}^{3}/{3}+c u_{y}^{2}/{2}$ and the elasticity along the
$y$-direction can be neglected if $j_{c}-j\ll j_{c},$ i.~e., the
problem can be reduced to the 1D massive string problem.  Above we
have generalized the problem for the case of an arbitrary potential
and nonzero elasticity in the $y$-direction.

\section{Decay rate}
\label{dr}

The decay rate $\Gamma$ of an arbitrary metastable Hamiltonian system
at high temperatures is given by the expression\cite{Langer1967}
\begin{equation}
\hbar\Gamma=\frac{\hbar{\omega}_{0}}{\pi} \frac{{\rm Im }Z}{Z},
\label{decay}
\end{equation}
where ${\omega}_{0}$ is the unstable mode growth rate and $Z$ is the
partition function of the system under consideration.
Eq.~(\ref{decay}) is applicable if the temperature $T$ satisfies the
condition $T\gg \hbar{\omega}_{0}$, otherwise quantum effects become
relevant.  If the transition from quantum to classical behavior in the
decay of the metastable state is of second order, the parameter
$T_{0}={\hbar{\omega}_{0}}/{2\pi}$ determines the temperature of the
crossover.  However, in general the transition can be of first order,
in which case the crossover temperature differs from
${\hbar{\omega}_{0}}/{2\pi}$ (see Ref.\cite{Lifshitz}).  Note that at
low temperatures the equation for the decay rate can be written as
$\hbar\Gamma =2T\thinspace {{\rm Im}Z}/{Z}.$ We emphasize that
Eq.~(\ref{decay}) can be applied only if the system is properly
equilibrized, i.~e., if its characteristic relaxation time is much
smaller than $\Gamma^{-1}.$

We assume that the quasiclassical approximation is applicable.  With
the partition function written as a path integral,
\begin{equation}
Z=\int\{{\cal D}{\bf u}\}\exp{\left (-\frac{S_{Eucl}[{\bf u}]}
{\hbar}\right )},
\label{pth}
\end{equation}
we can use the steepest descent method for its calculation.  In this
approximation the partition function is determined by its stationary
points.  The most significant contribution to $Z$ arises from the
trajectory ${\bf u}(z,t)={\bf 0} $ --- the position of the minimum of
the potential, whereas the imaginary part of the partition function is
determined by the neighborhood of the saddle-point solution ${\bf
  u}(z,t)={\bf u}_{0}(z)$ which is time-independent at high
temperatures.  Hence, the problem is reduced to the calculation of the
unstable mode growth rate ${\omega}_{0},$ and of the real and
imaginary parts of the statistical sum.  Below we shall calculate all
these quantities.

\subsection{Unstable mode growth rate}
\label{un}

At high temperatures the saddle-point solution does not depend on
time. The stationary extremal trajectory ${\bf u}_{0}(z)$ for the
Euclidean action
\begin{equation}
S_{\rm Eucl}[u_{x},u_{y}]=\int_{{-L}/{2}}^{{+L}/{2}}dz
\int_{-{\hbar}/{2T}}^{+{\hbar}/{2T}}d\tau \left (G[u_{x},u_{y}]
-{i\alpha}u_{y}\frac{\partial u_{x}}{\partial \tau}\right )
\label{Eact}
\end{equation}
corresponding to the real-time Lagrangian (\ref{real}) satisfies the
Euler equation
\begin{equation}
\epsilon
\frac{{\partial}^{2}{\bf u}}{\partial z^{2}}=
\frac{{\partial}U({\bf u})}{\partial {\bf u}},\ \
{\bf u}=(u_{x}, u_{y}).
\label{Newton}
\end{equation}
The unstable mode growth rate then is determined by the negative
eigenvalue of the operator $\delta_{\bf u}^{2}S_{\rm Eucl}|_{{\bf
    u_{0}}(z)}.$ Near the saddle-point solution ${{\bf u}_{0}}(z)$ the
perturbed solution can be expanded in the form
\begin{equation}
u_{x}(z,\tau)=u_{x0}(z)+\psi (z)\exp {\left ({i2\pi T\tau}/{\hbar}
\right )},
\label{xdist}
\end{equation}
\begin{equation}
u_{y}(z,\tau)=\varphi (z)\exp {\left ({i2\pi T\tau}/{\hbar} \right )},
\label{ydist}
\end{equation}
with small distortion amplitudes $\psi (z)$ and $\varphi (z)$ for
$T\alt T_{0}.$ Substituting $(\ref{xdist})$ and $(\ref{ydist})$ into
the equations of motion for the Euclidean action (\ref{Eact}) and
expanding, we obtain the system of equations determining the crossover
temperature $T_{0}$
\begin{equation}
-\epsilon{{d}^{2}\psi\over{d}{z}^{2}}
+\alpha{{\omega}_{0}}\varphi
+{{d}^{2}V_{cyl}\over{d}{r}^{2}}\bigg |_{r=u_{x0}(z)}
\equiv{\hat H}_{x}\psi+\alpha{{\omega}_{0}}\varphi=0,
\label{12}
\end{equation}
\begin{equation}
-\epsilon\frac{{d}^{2}\varphi}{{d}{z}^{2}}
-\alpha{{\omega}_{0}}\psi
+\frac{1}{r}{{d}V_{cyl}\over{d}{r}}\bigg |_{r=u_{x0}(z)}
\equiv{\hat H}_{y}\varphi-\alpha{{\omega}_{0}}\psi=0,
\label{13}
\end{equation}
subject to the boundary conditions $\psi ({\pm L}/{2}),\ \varphi ({\pm
  L}/{2}) =0.$ The (lowest) eigenvalue ${\omega}_{0}$ is related to
the crossover temperature $T_{0}$ through the condition
${\omega}_{0}={2\pi T_{0}}/{\hbar}.$

The operator ${\hat H_{x}}$ has only one negative eigenvalue: it can
be easily seen that in the limit $L\rightarrow\infty$ the function
${du_{x0}}/{dz}$ is an eigenfunction of the operator $\hat {H_{x}}$
with zero eigenvalue.  Its derivative $du_{x0}/dz$ can be understood
as the ``velocity of a particle'' moving in the potential
$U(u_{x0},u_{y0})$ with the ``velocity" changing its sign once. Hence,
$du_{x0}/dz$ is an eigenfunction of the 1D Schr\"odinger operator
${\hat H}_{x}$ with one node and there must be another function
corresponding to the ground state ``wave function" with a negative
eigenvalue.  We conclude that the operator ${\hat H}_{x}$ has one
negative and one zero eigenvalue.  The operator ${\hat H}_{y}$ is
positive and has only positive eigenvalues. The "potentials"
for both the operators ${\hat H}_{x}$ and ${\hat H}_{y}$ are plotted
in Figs.~1 and 2.

\centerline{\epsfxsize=9cm \epsfbox{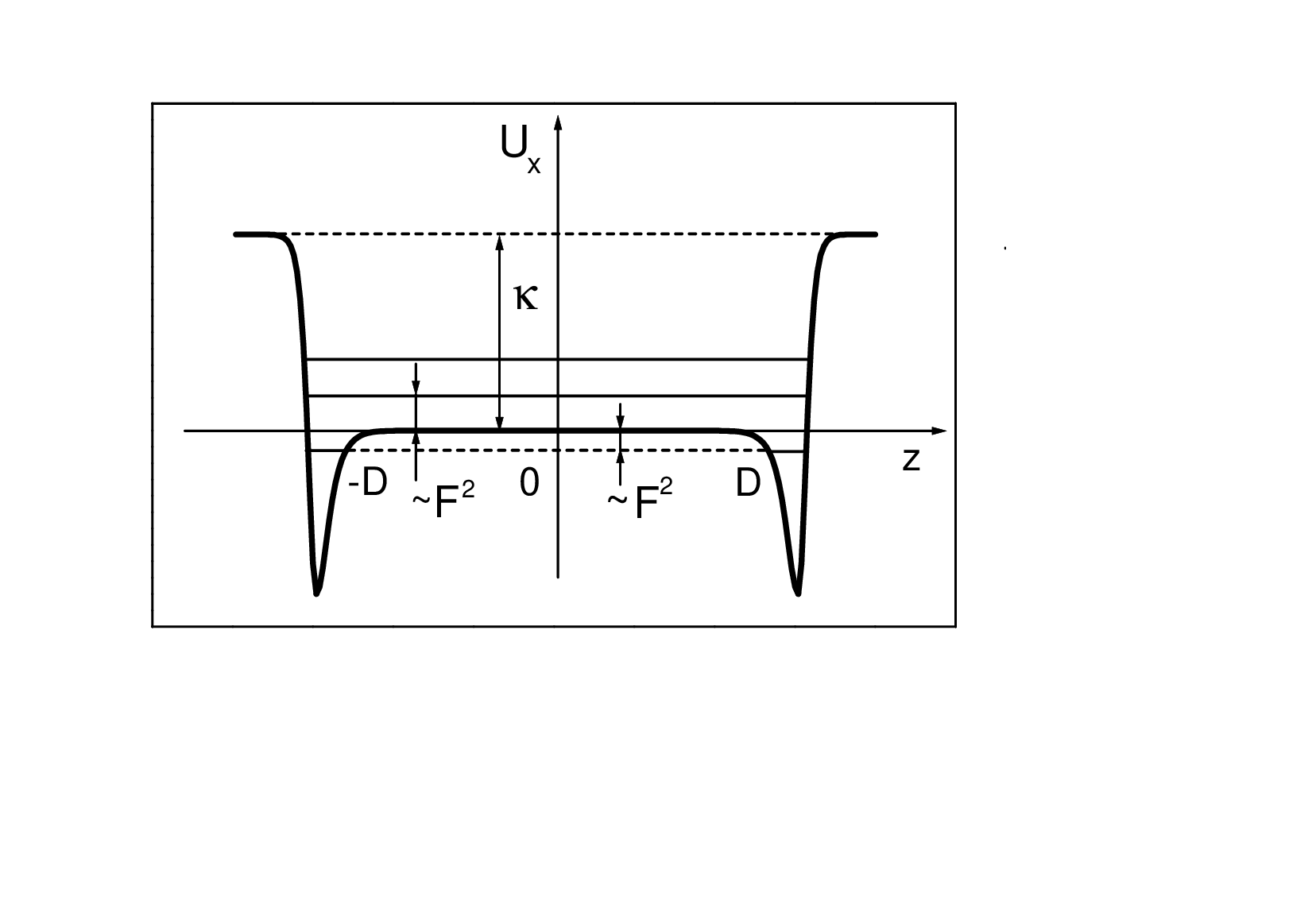}}
{\vskip-1.7cm{\footnotesize {\bf Fig.1}~The ``potential'' 
  $U_{x}(z)$ for the 1D Schr\"odinger operator ${\hat H}_{x}$ (see
  Eq.~(\ref{12})).  The characteristic size of the potential is given
  by $D={\sqrt{2\epsilon V_{0}}}/{F}$ and its height is equal
  to~$\kappa.$}}

\centerline{\epsfxsize=9cm \epsfbox{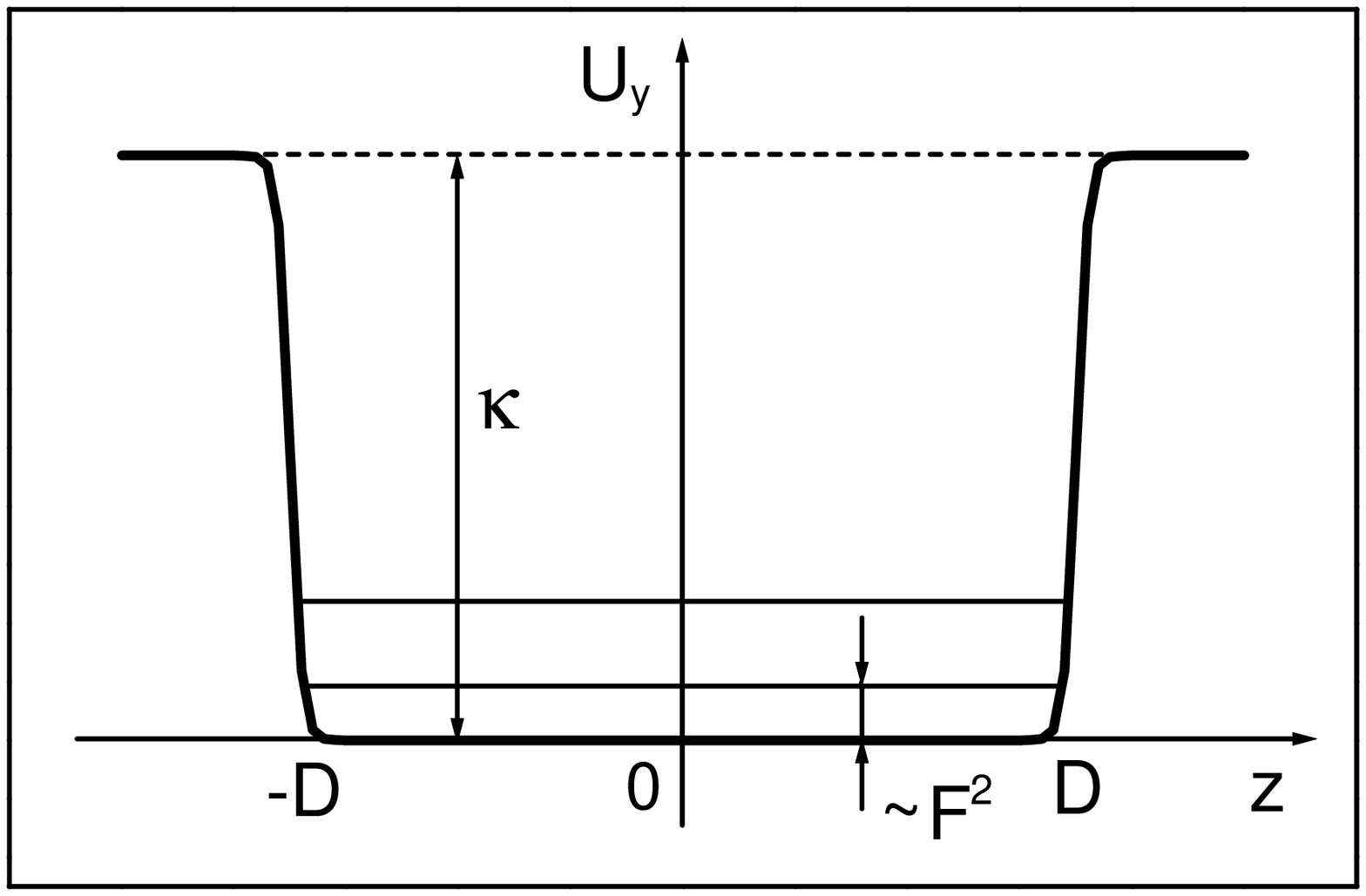}}
{\vskip-1.7cm{\footnotesize {\bf Fig.2}~The ``potential'' 
  $U_{y}(z)$ for the 1D Schr\"odinger operator ${\hat H}_{y}$ (see
  Eq.~(\ref{13})).  The characteristic size of the potential is given
  by $D={\sqrt{2\epsilon V_{0}}}/{F}$ and its height is equal to~
  $\kappa.$}}
\vskip0.2cm

Eliminating the function $\varphi$ from the system of Eqs.~(\ref{12})
and (\ref{13}) we obtain (we denote eigenvalues of ${\hat H}_{x}{\hat
  H}_{y}$ by $\tilde \lambda$ and those of ${\hat H}_{x}$ by
$\lambda$)
\begin{equation}
{\hat H}_{y} {\hat H}_{x}\psi = {\tilde \lambda}\psi.
\label{Hall}
\end{equation}
The unstable mode growth rate $\omega_{0}$ is determined by the
negative eigenvalue ${\tilde \lambda}_{-1}$ of Eq.~(\ref{Hall}),
${\tilde\lambda}_{-1}=-{\left (\alpha\omega_{0}\right )}^{2}.$ In
Appendix~\ref{existence} we derive a variational principle for the
present problem and show that the negative eigenvalue of
Eq.~(\ref{Hall}) indeed exists and is unique. (We exploit that the
operator ${\hat H}_{y}^{{1}/{2}}{\hat H}_{x}{\hat H}_{y}^{{1}/{2}}$ is
Hermitian and has the same eigenvalues as ${\hat H}_{y}{\hat H}_{x}$).
Furthermore, we derive both upper and lower bounds on the eigenvalue
${\tilde \lambda}_{-1}$ of the operator ${\hat H}_{x}{\hat H}_{y}$ in
Appendix~\ref{loww}, such that we finally arrive at the estimate
\begin{equation}
T_{0}=
\frac{\hbar\sqrt{|{\tilde \lambda}_{-1}|}}{2\pi\alpha}=
\frac{\xi}{2\pi}
\frac{\hbar F^{2}}{\alpha V_{0}},
\label{ottwett}
\end{equation}
with $\xi\in [1.1530,2.7314]$ a universal constant and $V_{0}$ denotes
the depth of the potential $V_{cyl}.$ The result $(\ref{ottwett})$
holds independently of the details of the pinning potential $V_{cyl}$
as long as the driving force $F$ is small, $F\ll F_{c}.$ We note that
$T_{0}\sim F^{2};$ the same dependence holds for a string with
dissipative dynamics.  For a massive string
depinning from a linear defect $T_{0}\sim F,$ see
Ref.\cite{Skvortsov}.

\subsection{Ratio ${{\rm Im}Z}/{Z}$}

The most significant contribution to ${\rm Re} Z$ arises from the
neighborhood of the minimum of the potential $V_{cyl}.$ In this region
we can write (see Eq.~(\ref{pott}))
\begin{equation}
U(x,{p}/{\alpha})\cong\frac{\kappa}{2}x^{2}+
\frac{\kappa}{2{\alpha}^{2}}p^{2},
\end{equation}
where $\kappa ={d^{2}V_{cyl}}/{du_{x}^{2}}|_{{\bf u}=0}.$ The
equations of motion (\ref{q1}) and (\ref{q2}) take the form
\begin{equation}
{\alpha}^{2}{\dot x}=\left (\kappa-
\epsilon\frac{{\partial}^{2}}{\partial z^{2}}\right )p
\equiv
{{\hat {H}}_{0}}p,
\label{H0}
\end{equation}
\begin{equation}
{\dot p}=-{\hat {H}}_{0}x.
\end{equation}
Hence, $p={\alpha}^{2}{\hat H}_{0}^{-1}{\dot x}$ and the Euclidean
action of the vortex near the equilibrium position can be written in
the form (we make use of Eq.~(\ref{hamm}))
\begin{equation}
S_{\rm Eucl}^{0}=
\frac{1}{2}\int_{-{\hbar}/{2T}}^{+{\hbar}/{2T}}d\tau
\int_{{-L}/{2}}^{{+L}/{2}}dz\thinspace x\left (-{\alpha}^{2}
\frac{{\partial}^{2}}{\partial {\tau}^{2}}{\hat H}_{0}^{-1}+
{\hat H}_{0}\right )x.
\label{rl}
\end{equation}
The same procedure as above provides the variation of the Euclidean
action near the thermal saddle-point solution
\begin{eqnarray}
S_{\rm Eucl}^{S}=
\frac{1}{2}\int_{-{\hbar}/{2T}}^{+{\hbar}/{2T}}d\tau
\int_{{-L}/{2}}^{{+L}/{2}}dz\thinspace
\delta x\left (-{\alpha}^{2}\frac{{\partial}^{2}}
{\partial {\tau}^{2}}{\hat H}_{y}^{-1}
+{\hat H}_{x} \right )\delta x\thinspace,
\label{imag}
\end{eqnarray}
where the operators ${\hat H}_{x}$ and ${\hat H}_{y}$ are given by
Eqs.~(\ref{12}) and (\ref{13}).  Using the usual measures for harmonic
oscillators in the path integrals for $S_{\rm Eucl}^{0}$ and $S_{\rm
  Eucl}^{S}$ (see Ref.\cite{Zinn-Justin}),
\begin{equation}
d\mu^{0}={\left [{\rm det} \left (-{\alpha}^{2}\frac{\partial^{2}}
{\partial {\tau}^{2}}
{\hat H}_{0}^{{-1}}
\right )\right ]}^{{1}/{2}}
\prod_{a}\frac{dC_{a}^{0}}{\sqrt{2\pi\hbar}}
\end{equation}
and 
\begin{equation}
d{\mu^{S}}=
{\left [{\rm det} \left (-{\alpha}^{2}\frac{\partial^{2}}
{\partial \tau^{2}}
{\hat H}_{y}^{{-1}}
\right )\right ]}^{{1}/{2}}
\prod_{b}\frac{dC_{b}^{S}}{\sqrt{2\pi\hbar}},
\end{equation}
(here $C_{a}^{0}$ and $C_{b}^{S}$ are expansion coefficients over the
systems of normalized functions) and defining the dimensionless
operators ${\hat h}_{0},$ ${\hat h}_{x},$ and ${\hat h}_{y}$ as
${{\hat H}_{0}}/{\kappa},$ ${{\hat H}_{x}}/{\kappa},$ and ${{\hat
    H}_{y}}/{\kappa},$ respectively
($\kappa=V_{cyl}^{\prime\prime}\left (0\right )$), we obtain for
${{\rm Im}Z}/{Z}$
\begin{eqnarray}
\frac{{\rm Im}Z}{Z}=
& \thinspace &
\frac{L}{2}
\sqrt{\frac{\kappa}{{2\pi T}}}
{\left [\int_{-{L}/{2}}^{+{L}/{2}}dz
{\left (\frac{\partial u_{0}}{\partial z} \right ) }^{2}
\right ]}^{{1}/{2}}
\exp{\left (-\frac{U}{T}\right )}
\nonumber\\
& \thinspace &
{\left [
\frac
{\mu_{x,0}
({L})}
{\mu_{xy,0}({L})}
\right ]}^{{1}/{2}}
{\left |
\frac
{{{\rm det}
{\left
(-\frac{{\alpha}^{2}}{\kappa^{2}}
\frac{{\partial}^{2}}{\partial \tau^{2}}+{{\hat h}_{0}}^{2}
\right )}
}}
{{{{\rm det}^{\prime}}
{\left
(-\frac{{\alpha}^{2}}{{\kappa}^{2}}
\frac{{\partial}^{2}}{\partial \tau^{2}}+
{{\hat h}_{y}}^{{1}/{2}}{\hat h}_{x}
{{\hat h}_{y}}^{{1}/{2}}
\right )}
    }}
\right |}
^{{1}/{2}},
\label{dZnew}
\end{eqnarray}
where ($L\rightarrow\infty $)
\begin{equation}
U=\epsilon\int\limits_{{-L}/{2}}^{{+L}/{2}}
{\left (\frac{\partial {{u}_{0}}}{\partial z}\right )}^{2}dz\simeq
\frac{4}{3}\sqrt{2\epsilon V_{0}}\frac{V_{0}}{F},\ \ \ 
\label{barrier}
\end{equation}
is the activation barrier and $L$ is the length of the string (we
performed standard integration over the shift mode, see
Refs.\cite{Rajaraman,Ivlev}).  The prime in Eq.~(\ref{dZnew})
indicates that we exclude the zero eigenvalue of the operator
$-{\alpha}^{2}{{\hat H}_{y}^{-1}} \frac{{\partial}^{2}}{\partial
  \tau^{2}}+{\hat H}_{x}.$ $\mu_{x,0}\left ({L}\right )$ and
$\mu_{xy,0}\left ( L \right )$ are the ``zero'' eigenvalues of the
operators ${\hat h}_{x}$ and ${\hat h}_{y}^{{1}/{2}}{\hat h}_{x}{\hat
  h}_{y}^{{1}/{2}}$ defined on the interval $[-{ L}/{2},\thinspace +{
  L}/{2}],$ $L\rightarrow\infty.$

The eigenvalues of the operators $-\frac{{\alpha}^{2}}{\kappa^{2}}
\frac{{\partial}^{2}}{\partial \tau^{2}}+{{\hat h}_{0}}^{2}$ and
$-\frac{{\alpha}^{2}}{{\kappa}^{2}} \frac{{\partial}^{2}}{\partial
  \tau^{2}}+ {{\hat h}_{y}}^{{1}/{2}}{\hat h}_{x} {{\hat
    h}_{y}}^{{1}/{2}}$ are equal to
\begin{equation}
{\left (\frac{2\pi\alpha T}{\kappa\hbar}n\right )}^{2}+
\mu_{0,a}\ \ \ \
{\rm and}\ \ \ \
{\left (\frac{2\pi\alpha T}{\kappa\hbar}n\right )}^{2}+
\mu_{xy,b},\ \ \ n=0,\pm 1,\pm 2\dots ,
\end{equation}
with $\mu_{0,a}$ and $\mu_{xy,b}$ the eigenvalues of the operators
${\hat h}_{0}^{2}$ and ${{\hat h}_{y}}^{{1}/{2}}{\hat h}_{x} {{\hat
    h}_{y}}^{{1}/{2}},$ respectively.  Substituting these eigenvalues
into Eq.~(\ref{dZnew}) and calculating the product over $n$ using
$\prod_{n=1}^{\infty}\left (1+{x^{2}}/{n^{2}}\right )= {\sinh{(\pi x
    )}}/{\pi x}$ we obtain
\begin{eqnarray}
& \thinspace &
\frac{{\rm Im}Z}{Z}=
\frac{\alpha L}{4\hbar}
\sqrt{\frac{T}{{2\pi\kappa}}}\exp{\left (-\frac{U}{T}\right )}
\frac{1}{\sin{\frac{\pi T_{0}}{T}}}
{\left [
\frac
{\mu_{x,0}
\left ({ L}\right )}
{\mu_{xy,0}\left ({L}\right )}
\right ]}^{{1}/{2}}
\nonumber\\
& \thinspace &
{\left [\int_{{-L}/{2}}^{{+L}/{2}}dz
{\left (\frac{\partial u_{0}}{\partial z} \right )}^{2}
\right ]}^{{1}/{2}}
\exp{\left [
\frac{\hbar\kappa}{2\alpha T}
\left (
{\sum\limits_{a}}\sqrt{\mu_{0,a}}
-{\sum\limits_{b}}^{\prime\prime}
\sqrt{{\mu}_{xy,b}}
\right )
\right ]}
\nonumber\\
& \thinspace &
\frac{
{\prod\limits_{a}}
\left [1-\exp{
\left (-\frac{\hbar\kappa}{\alpha T}
\sqrt{\mu_{0,a}}\right )  }\right ]    }
{
{\prod\limits_{b}}^{\prime\prime}
\left [1-\exp{\left (-\frac{\hbar\kappa}{\alpha T}
\sqrt{{ \mu}_{xy,b}}\right )}\right ]  
}.
\label{findz}
\end{eqnarray}
The double-prime sign reminds us that we excluded the negative and
zero eigenvalues of the operator ${{\hat h}_{y}}^{{1}/{2}} {\hat
  h}_{x} {{\hat h}_{y}}^{{1}/{2}}.$

\subsection{Result for $\Gamma$}

The positive part of the spectrum of ${\hat h}_{x}^{{1}/{2}}{\hat
  h}_{y}{\hat h}_{x}^{{1}/{2}}$ consists of $M$ discrete eigenvalues
$\mu_{xy,b}$ and a continuous part $\mu_{xy} (k)={\left (1+{\epsilon
      k^{2}}/{\kappa} \right )}^{2}$ with a spectral density
$\rho^{S}(k).$ The spectrum of ${\hat h}_{0}^{2}$ is continuous,
$\mu_{0} (k)={\left (1+{\epsilon k^{2}}/{\kappa} \right )}^{2}$ with
the spectral density $\rho^{0}(k).$ Using
\begin{equation}
\frac
{\prod\limits_{a}f(\mu_{0,a})}
{\prod\limits_{b}f(\mu_{xy,b})}=
\exp
\left [
{\int
 dk\thinspace \delta\rho (k)\ln f\left (\mu (k)\right )}
\right ]
\end{equation}
with $\delta \rho (k)=\rho^{0}(k)-\rho^{S}(k)$ we arrive at the final
expression for the decay rate.

\begin{eqnarray}
& \thinspace &
\Gamma =
\frac{\alpha T_{0}}{2\hbar^{2}}
\sqrt{\frac{T}{{2\pi\kappa}}}\exp{\left (-\frac{U^{*}}{T}\right )}
\frac{L}{\sin{\frac{\pi T_{0}}{T}}}
{\left [\int_{-{L}/{2}}^{+{L}/{2}}dz
{\left (\frac{\partial u_{0}}{\partial z} \right )}^{2}
\right ]}^{{1}/{2}}
{\left [
\frac
{\mu_{x,0}
\left ({L}\right )}
{\mu_{xy,0}
\left ({ L}\right )}
\right ]}^{{1}/{2}}
\nonumber\\
& \thinspace &
\exp{\left (
\int\limits_{0}^{\infty} dk\thinspace \delta\rho (k)
\ln{\left [
1-\exp{ \left (-\frac{\hbar \kappa}{\alpha T}
\left (1+\frac{\epsilon}{\kappa}k^{2}\right )
\right )}\right ] }\right )}\thinspace
{
{\prod\limits_{b=1}^{M}}
\left [
1-\exp{\left (-\frac{\hbar \kappa}{\alpha T}\sqrt{{\mu_{xy,b}}},
\right )}\right ]}^{-1},
\label{otw}
\end{eqnarray}
with
\begin{equation}
U^{*}=
{U}
-\frac
{\hbar \kappa}{2\alpha}
\left [
\int\limits_{0}^{\infty} dk\thinspace \delta\rho(k)
\left (1+\frac{\epsilon}{\kappa}k^{2}\right )
-{\sum\limits_{b=1}^{M}}{\sqrt{{\mu_{xy,b}}}}
\right ]
\label{renormnew}
\end{equation}
the quantum renormalized activation energy.  Let us investigate the
correction to the activation energy (\ref{renormnew}) arising from
large $k$-values.  We will see that the large $k$ modes lead to an
ultraviolet divergence which we have to cut off by some physical
length scale.  It is convenient to rewrite the integral
$\int\limits_{0}^{k} dk^{\prime}\thinspace \delta\rho(k^{\prime})
\left (1+{{\epsilon}{k^{\prime}}^{2}}/{\kappa}\right )$ as
\begin{equation}
\int\limits_{0}^{k} dk^{\prime}\thinspace \delta\rho(k^{\prime})
\left (1+\frac{\epsilon}{\kappa}{k^{\prime}}^{2}\right )=
\sum\limits_{n}^{N_{0}(k)}\left (1+{\epsilon k_{n}^{2}}/{\kappa^{2}}\right )-
\sum\limits_{n}^{N_{S}(k)}\left (1+{\epsilon q_{n}^{2}}/{\kappa^{2}}\right ),
\end{equation}
where $k_{n}$ and $q_{n}$ are trivially related to the eigenvalues of
the operators ${\hat h}_{0}^{2}$ and ${\hat h}_{y}^{{1}/{2}}{\hat
  h}_{x} {\hat h}_{y}^{{1}/{2}}$ and $N_{0}(k)$ and $N_{S}(k)$ denote
the number of eigenvalues inside the interval $[0,k].$ Note that
$N_{0}(\infty )-N_{S}(\infty )=M+2.$ The two operators ${\hat
  h}^{2}_{0}$ and ${\hat h}_{y}^{{1}/{2}}{\hat h}_{x}{\hat
  h}_{y}^{{1}/{2}}$ define two scattering problems. We define
appropriate scattering states $\psi_{k}(z)$ and $\psi_{q}(z)$ with
asymptotics $e^{ikz}$ and $e^{iqz}$ ($z\rightarrow -\infty $) and
$e^{ikz},e^{iqz+\delta(q)}$ ($z\rightarrow +\infty $), with $\delta
(q)$ the appropriate scattering phase shift.  The general solutions of
the ``Schr\"odinger'' equations can be written as
$\Psi_{k}(z)=C_{1}\psi_{k}(z)+ C_{2}\psi_{-k}(z)$ and
$\Psi_{q}(z)=C_{1}^{\prime}\psi_{q}(z)+ C_{2}^{\prime}\psi_{-q}(z).$
Since we require that the ``wave functions'' satisfy the condition
$\Psi_{p}(\pm {L}/{2})=0,$ ${L}\gg {\sqrt{\epsilon V_{0}}}/{F},$
we obtain the discrete levels through the phase equations
(see also Ref.\cite{Vainshtein})
\begin{equation}
k_{n}{L}=\pi n,\ \ \ \ \ q_{n}{L}+\delta (q)=\pi n ,
\end{equation}
and hence
\begin{equation}
k_{n}^{2}-q_{n}^{2}=
{\left (\frac{\pi n}{L}\right )}^{2}-
{\left (\frac{\pi n-\delta (q)}{L}\right )}^{2}
\simeq
+\frac{2\pi n}{{L}^{2}}\delta (q)
\simeq
+\frac{2q}{L}\delta (q).
\end{equation}
For $q\gg\sqrt{{\kappa}/{\epsilon}}$ we can make use of the
semiclassical approximation for the operator ${\hat
  h}_{y}^{{1}/{2}}{\hat h}_{x}{\hat h}_{y}^{{1}/{2}}$ and calculate
the phase shift directly: With the usual Ansatz $\Psi (z)\sim e^{i\int
  q(z)\thinspace dz}$ the function $q(z)$ can be found from the
equation
\begin{equation}
\left (
\epsilon q^{2}(z)+U_{x}
\right )
\left (
\epsilon q^{2}(z)+U_{y}
\right )={\kappa}^{2}E_{q},
\end{equation}
where $E_{q}= {\left (1+{\epsilon q^{2}}/{\kappa}\right )}^{2} \simeq
{\epsilon^{2}q^{4}}/{\kappa^{2}}$ is the eigenvalue of the continuous
spectrum of ${\hat h}_{y}^{{1}/{2}}{\hat h}_{x}^{{1}/{2}}{\hat
  h}_{y}^{{1}/{2}}.$ At large $q$ we obtain
\begin{equation}
q(z)\simeq
\sqrt{\frac{\kappa}{\epsilon}}E_{q}^{{1}/{4}}
\left (1-\frac{(U_{x}+U_{y})}{4\kappa\sqrt{E_{q}}}\right ).
\end{equation}
Similarly,
for the operator ${\hat h}_{0}^{2}$ we obtain
\begin{equation}
k(z)\simeq
\sqrt{\frac{\kappa}{\epsilon}}E_{k}^{{1}/{4}}
\left (1-\frac{1}{2\sqrt{E_{k}}}\right ),
\end{equation}
and we arrive at the difference in the phase shift
\begin{equation}
\delta (q)\simeq\sqrt{\frac{\kappa}{\epsilon}}E_{q}^{{1}/{4}}
\int\limits_{-{L}/{2}}^{+{L}/{2}}dz
\frac{\left (2\kappa -U_{x} -U_{y}\right )}
{4\kappa\sqrt{E_{q}}}.
\end{equation}
Taking into account that at large $q,$ $E_{q}\simeq
{\epsilon^{2}q^{4}}/{\kappa^{2}}$ and that $U_{x}(z),U_{y}(z)\ll
\kappa$ for $|z|<{\sqrt{2\epsilon V_{0}}}/{F},$ we obtain
\begin{equation}
\delta ({q})\simeq
\sqrt{\frac{2V_{0}}{\epsilon}}
\frac{\kappa}{qF},
\end{equation}
i.e., $k_{n}^{2}-q_{n}^{2}= 2\kappa\sqrt{{2V_{0}}/{\epsilon}}/{FL}.$
At large $k$ the number of states in the interval $dk$ is equal to ${L
  dk}/{2\pi}$ and the integral
\begin{equation}
\int\limits_{0}^{k}dk^{\prime}\delta\rho (k^{\prime})\left (1+
\frac{\epsilon {k^{\prime}}^{2}}{\kappa} \right )\simeq
\frac{\epsilon}{\kappa}
\sum\limits_{n}^{N_{0}(k)}\left (k_{n}^{2}-q_{n}^{2}\right )
\simeq\int\limits_{0}^{k} dk^{\prime}
\frac{\sqrt{{2\epsilon V_{0}}}}{\pi F}
\end{equation}
diverges linearly at large $k$ (for the case of a massive string the
correction to the activation diverges as $\ln{k},$ see
Refs.\cite{Skvortsov,Ivlev}).  The integral then has to be cut off at
some wave vector $k^{*}:$ For a string the natural cutoff is
${\pi}/{r},$ with $r$ the radius of the string. For a vortex parallel
to the $c$-axis of an anisotropic superconductor ${k^{*}}\sim{\pi}/
{\max \left (\xi_{c},d\right )},$ with $\xi_{c}$ the coherence length
in the $c$-direction and $d$ the distance between the superconducting
layers.

Cutting off the integral in (\ref{renormnew}) at large momentum
$k^{*}$ we obtain
\begin{equation}
U^{*}\simeq U-\frac{\hbar\kappa k^{*}}{\sqrt{2}\pi\alpha F}
\sqrt{{V_{0}}{\epsilon}}.
\label{explren}
\end{equation}
The theory is self-consistent if the correction to the activation
energy is small as compared to $U$ (here we have assumed the cutoff
$k^{*}$ to be large such that we can neglect the contribution from the
bound states).

In case of a continuous transition from quantum to classical behavior
the result Eq.~(\ref{otw}) with Eq.~(\ref{renormnew}) is applicable
for any $T>T_{0}$ except for a narrow temperature interval $\sim
{\hbar}^{{3}/{2}}$ around $T_{0}$ (see Ref.\cite{Affleck}).  An
explicit expression for the decay rate can be obtained only for the
case $\hbar\rightarrow 0,$ i.~e., in the classical limit.  In this
limit we can expand the exponents in (\ref{otw}), $1- \exp{\left
    (-{\hbar\kappa}/{\alpha T}\right )}\simeq {\hbar\kappa}/{\alpha
  T}$ and the quantum correction to the activation energy
(\ref{renormnew}) tends to zero.  Performing this steps we obtain the
result
\begin{eqnarray}
\Gamma=
\frac{\kappa L}{2\pi\alpha}\sqrt{\frac{\kappa}{2\pi T}}
\exp{\left (-\frac{U}{T}
\right )}
{\left [\int_{-{L}/{2}}^{+{L}/{2}}dz
{\left (\frac{\partial u_{0}}{\partial z} \right )}^{2}
\right ]}^{{1}/{2}}
{\left [
\frac
{\mu_{x,0}
\left ({L}\right )}
{\mu_{xy,0}
\left ({L}\right )}
\right ]}^{{1}/{2}}
{\left [
\frac{{{\rm det} \left ({{\hat h}_{0}^{2}}\right )}}
{{\rm det}^{\prime\prime}
\left (
{\hat h}_{y}^{{1}/{2}}
{{\hat h}_{x}}
{\hat h}_{y}^{{1}/{2}}
\right )}
\right ]}
^{{1}/{2}}.
\end{eqnarray}
Using the same technique as before, see Eq.~(\ref{dZnew}), we arrive
at the more suitable form
\begin{eqnarray}
&\thinspace &
\Gamma = 
\frac{\kappa L}{2\pi\alpha}\sqrt{\frac{\kappa}{2\pi T}}
\exp{\left (-\frac{U}{T}
\right )}
{\left [\int_{-{L}/{2}}^{+{L}/{2}}dz
{\left (\frac{\partial u_{0}}{\partial z} \right )}^{2}
\right ]}^{{1}/{2}}
\nonumber\\
&\thinspace  &
\sqrt{|{\mu}_{xy,-1}|}
\lim_{{L}\rightarrow \infty}
\sqrt{ \mu_{x,0}({L})}
\thinspace
{\left |
\frac{{{\rm det} \left ({{\hat h}_{0}}\right )}}
{{\rm det}
\left (
{{\hat h}_{x}}{\left ({L}\right )}
\right )}
\right |}
^{{1}/{2}}
{\left [
\frac{{{\rm det} \left ({{\hat h}_{0}}\right )}}
{{\rm det}
\left (
{{\hat h}_{y}}
\right )}
\right ]}
^{{1}/{2}},
\label{final22}
\end{eqnarray}
where $\mu_{x,0}({L})$ is the zero eigenvalue of the operator ${\hat
  h}_{x}(L).$

The calculation of the instanton determinant ratios (\ref{final22})
can be elegantly performed using the Gelfand-Yaglom
formula\cite{Gelfand}.  The renormalized\cite{Gelfand} determinant
${\rm det}{\hat H}$ of the Schr\"odinger operator ${\hat H}=
-{\partial^{2}}/{\partial z^{2}}+p(z)$ is the solution of the
differential equation ${\hat H}{ f{\left (z\right )}}=0$ with the
initial conditions ${f}|_{z=-{L}/{2}} =0$ and
${df}/{dz}|_{z=-{L}/{2}}=1.$ The value of the function
${f}(z={L}/{2})$ provides the renormalized determinant.  As has been
shown in Ref.\cite{Kramer1}, in the limit $F\rightarrow 0,$
$\mu_{x,0}\left (L\right )\sim \left ({Fa}/{V_{0}}\right ) \exp{\left
    [-\sqrt{{\kappa}/{\epsilon}}\left (L-{2\sqrt{{2\epsilon
            V_{0}}}/{F}} \right )\right ]},$ ${{\rm det}({\hat
    h}_{0})}/ |{\rm det} ({\hat h}_{x}(L) )|\sim \exp{\left
    (\sqrt{{\kappa}/{\epsilon}}\thinspace L\right )},$ and ${{\rm
    det}({\hat h}_{0})}/ {{\rm det} ({\hat h}_{y} )}\sim \left
  ({Fa}/{V_{0}}\right ) \exp{\left ({2\sqrt{2\kappa V_{0}}}/{F}\right
  )},$ where the coefficients of proportionality depend on the
detailed form of the pinning potential and $a$ is the characteristic
radius of the pinning well. The eigenvalue
$\mu_{xy,-1}=-{\xi^{2}F^{4}}/{\kappa^{2} V_{0}^{2}},$ see
Eq.~(\ref{ottwett}), and we make use of Eq.~(\ref{barrier}).
Substituting these dependences into Eq.~(\ref{final22}) we obtain
\begin{equation}
\Gamma =
g\frac{V_{0}L}{\alpha a^{3}}
\sqrt{\frac{V_{0}}{\epsilon}}
{\left (\frac{F}{F_{c}}\right )}^{{5}/{2}}
{\left (\frac{\tilde T}{T}\right )}^{{1}/{2}}
\exp{\left (-\frac{U}{T}+\frac{2\sqrt{2\kappa V_{0}}}{F} \right )},
\label{highhigh}
\end{equation}
where $g$ is a dimensionless proportionality coefficient which depends
on the detailed form of the pinning potential, ${\tilde
  T}=a\sqrt{\epsilon V_{0}},$ $F_{c}$ is the critical force, and the
activation energy $U$ is given by Eq.~(\ref{barrier}).  As an
illustration we have explicitly solved the differential equations for
a pinning potential given by the equation
\begin{equation}
V_{cyl}=V_{0}
\left [1-
\cos^{2}{\left( \frac{\pi}{2a}\sqrt{u_{x}^{2}+u_{y}^{2}}
\right )}\theta\left (a^{2}-u_{x}^{2}-u_{y}^{2}\right )\right ],
\label{toy}
\end{equation}
where $\theta (t)$ is the step-function ($\theta (t)=1$ if $t>0$ and
$\theta (t)=0$ otherwise).  The external force has the form
$F(u_{x})=F\theta (|u_{x}|-a).$

We find $\mu_{x,0}\left (L\right )=\left ({24Fa}/{\pi V_{0}}\right
)\exp {\left [-\sqrt{{\kappa}/{\epsilon}} \left (L-{2\sqrt{{2\epsilon
            V_{0}}}/{F}} \right )\right ]},$ ${{\rm det}({\hat
    h}_{0})}/ |{\rm det} ({\hat h}_{x} )|=({1}/{2}) \exp{\left
    (\sqrt{{\kappa}/{\epsilon}}\thinspace L\right )},$ and ${{\rm
    det}({\hat h}_{0})}/ {{\rm det} ({\hat h}_{y} )}= {4Fa}/{\pi
  V_{0}} \exp{\left ({2\sqrt{2\kappa V_{0}}}/{F}\right )},$ and
substituting these expressions into Eq.~(\ref{final22}) we obtain
$g={\pi\xi}/{2^{{5}/{4}}}\simeq 1.32\xi.$ 

A real physical system will
involve a finite cutoff $k^{*},$ which we can account for in
Eq.~(\ref{otw}) by a restricted integration over the modes $k<k^{*}.$
However, in making use of the Gelfand-Yaglom formula we actually
account for unphysical fluctuations with $k>k^{*}.$ In order for the
classical result (\ref{highhigh}) to be valid we then must require the
cutoff $k^{*}$ to be sufficiently large such as to validate the
Gelfand-Yaglom approach.  On the other hand, $k^{*}$ has to be small
enough in order to arrive at a small quantum correction in
Eq.~(\ref{renormnew}).

\section{Conclusion}
\label{con}

Let us first specify the regime of applicability of our results.  The
zero temperature Euclidean action takes the value $S_{\rm Eucl}\left
  (T=0\right )=\alpha \Omega,$ where $\Omega$ is the volume encircled
by the string in the course of the tunneling motion. In the $x$- and
$y$-directions the characteristic size of the loop is ${V_{0}}/{F}$
and the length of the string segment along the $z$-direction is of
order ${\sqrt{\epsilon V_{0}}}/{F},$ hence $S_{\rm Eucl} \left (T=0
\right ) \sim \alpha \sqrt{{\epsilon}/{V_{0}}} {\left
    ({V_{0}}/{F}\right )}^{3}.$ On the other hand, $S_{\rm Eucl}\left
  (T=T_{0}\right )={\hbar U}/{T_{0}}\sim S_{\rm Eucl} \left (T=0
\right ),$ i.e., even if the transition from quantum to classical
behavior at $T_{c}> T_{0}$ is first order like, its temperature will
still be of order $T_{c}\sim T_{0}$ and our results are applicable in
the regime $T>T_{0}.$

Next, we estimate the temperature $T_{0}$ using parameters for the
moderately anisotropic superonductor ${\rm
  YBa_{2}Cu_{3}O_{7-\delta}}.$ We choose our columnar defect in the
form of a cylindrical cavity of radius $a\simeq\xi_{ab}(0),$ where
$\xi_{ab}$ is the coherence length in the $ab$-plane.  The expression
for the pinning potential then can be written in the form\cite{Nelson}
\begin{equation}
U(r)=\nu\frac{\epsilon_{0}}{4}\frac{r^{2}}{r^{2}+2\xi_{ab}^{2}},
\label{nelvin}
\end{equation}
where $\epsilon_{0}={\left ({\Phi_{0}}/{4\pi\lambda_{ab}}\right
  )}^{2}$ is the relevant energy scale and the factor $\nu$ ($0<\nu
<1$) describes the pinning efficiency factor\cite{Krusin}, $\Phi_{0}$
and $\lambda_{ab}$ are the flux quantum and the penetration length in
the $ab$-plane, respectively.  For the model potential (\ref{nelvin})
the temperature $T_{0}$, the critical force $F_{c},$ and the
temperature ${\hbar\kappa}/{\alpha}$ can be easily calculated (the
Hall coefficient $\alpha$ is related to the electron density $n$ as
$\alpha =\pi\hbar n$ at $T=0$ in the superclean limit
$\omega_{0}\tau\gg 1$): $T_{0}= \left ({\nu\xi}/{64\pi^{2}}\right )
\left ({\epsilon_{0}}/{n\xi_{ab}^{2}}\right ){\left ({F}/{F_{c}}\right
  )}^{2},$ $F_{c}=\left ({\sqrt{2}\nu}/{16}\right
){\epsilon_{0}}/{\xi_{ab}},$
${\hbar\kappa}/{\alpha}={\nu\hbar\epsilon_{0}}/{4\alpha\xi_{ab}^{2}}.$
Substituting $\xi_{ab}=16\thinspace {\rm \AA},$ 
$\lambda_{ab}=1400\thinspace{\rm \AA},$
${\lambda_{c}}/{\lambda_{ab}}=5,$ $\alpha=\pi \hbar n,$
$n=2\cdot 10^{21}\thinspace {\rm cm}^{-3},$
$\epsilon={\epsilon_{0}\lambda_{ab}^{2}}/{\lambda_{c}^{2}},$ and
$\xi\simeq2,$
we obtain
$T_{0}=0.6\thinspace {\rm K}\cdot\nu{\left ({F}/{F_{c}}\right )}^{2},$
and
${\hbar\kappa}/{\alpha}=15\thinspace {\rm K}\cdot \nu .$
Taking into account that $\nu$ might be $\sim 0.1,$ see Ref.\cite{Krusin},
we see that at $T\sim$10\thinspace--\thinspace15$\thinspace {\rm K}$
when the Hall force appears to be large\cite{Matsuda},
the conditions $T\gg T_{0},{\hbar\kappa}/{\alpha}$ are well satisfied.

The ratio ${\alpha}/{\eta}\simeq 15$ for $T\alt 15\thinspace{\rm K}$
(see Ref.\cite{Matsuda}) was obtained from indirect measurements.  In
Ref.\cite{Harris} direct measurements of the Hall angle in
60\thinspace{K} ${\rm YBa_{2}Cu_{3}O_{7-\delta}}$ are reported to
yield a ratio ${\alpha}/{\eta}\simeq 1.$ We wish to point out that
even if ${\alpha}/{\eta}\alt 1$ it is still possible to use
Eq.~(\ref{highhigh}): We have shown that the inverse lifetime of a
vortex pinned by a columnar defect behaves at high temperatures as
\begin{equation}
\Gamma\sim
F^{{5}/{2}}{T^{-{1}/{2}}}\exp{\left (-\frac{U}{T}+
\frac{2\sqrt{2\kappa V_{0}}}{F} \right )}.
\label{prop}
\end{equation}
It is interesting to note that the same dependence of the decay rate
on the external force and temperature has been obtained in
Refs.\cite{Kramer1} for a flux line governed by dissipative dynamics.
Moreover, the equation for the decay rate from Refs.\cite{Kramer1} can
be obtained up to a numerical prefactor from Eq.~(\ref{highhigh}) by
the substitution $\alpha\rightarrow\eta ,$ indicating that in the
regimes $\alpha\alt\eta$ or $\alpha\ll\eta$ it is possible to use
Eq.~(\ref{highhigh}) with $\alpha\rightarrow \max\left (\alpha,
  \eta\right ).$

The scaling law (\ref{prop}) for the decay rate leads then to the
resistivity scaling
\begin{equation}
\rho (F)\sim
\left [\sqrt{{\frac{F}{T}}}F
\exp{\left ({2\sqrt{2\kappa V_{0}}}/{F}\right )}\right]
\exp{\left(-\frac{U}{T}\right )},
\end{equation}
where the factor in square brackets represents the correction to the
standard result, which arises from the inclusion of classical
fluctuations around the saddle point (prefactor).

We note that Eq.~(\ref{prop}) is not valid for $T>{FU}/{2\sqrt{2\kappa
    V_{0}}}$ as the quasiclassical approximation is not applicable at
these large temperatures.  It is well-known that the problem of a
vortex pinned by a cylindrical potential is equivalent to that of a 2D
quantum particle trapped in a radially symmetric 2D potential.  The
latter always has a bound state in 2D, and thus the vortex is pinned
by the defect at any temperature. However, the thermal fluctuations
lead to a large downward renormalization of the pinning energy at
sufficiently high temperatures\cite{Nelson,Krusin}.

Finally, let us make an estimate of the cutoff momentum $k^{*}.$ The
theory developed above works well if
${k^{*}}\gg{\sqrt{{\kappa}/{\epsilon}}};$ with $k^{*}={\pi}/{d},$
where $d=12\thinspace{\rm \AA}$ is the distance between
superconducting layers, we obtain
${k^{*}}/{\sqrt{{\kappa}/{\epsilon}}}\simeq 5,$ such that this
condition is marginally satisfied.

Briefly summarizing, we have considered the problem of the thermally
activated depinning of a flux line governed by Hall dynamics from a
columnar defect in the presence of a small ($F\ll F_{c}$) external
force. We have shown how to reduce the 2D problem of the string motion
to a 1D effective problem.  The expression for the decay rate has been
obtained for the whole temperature region where the saddle-point
solution is time-independent (see Eq.~(\ref{otw}) and
Eqs.~(\ref{final22}) and (\ref{highhigh}) for high temperatures).  An
analytical expression for the classical asymptotics of $\Gamma$ has
been calculated for a model potential as given by Eq.~(\ref{toy}), and
we have discussed possible applications of these results to
high-$T_{c}$ superconductors.  \acknowledgements We gratefully
acknowledge financial support from the Swiss National Foundation.

\begin{appendix}
\section{Existence and Uniqueness}
\label{existence}

We prove that a solution of Eq.~(\ref{Hall}) with a negative
eigenvalue exists.  Second, we show the uniqueness of this solution.
We mention that in the limit $L\rightarrow\infty$ the boundary
condition $\psi\rightarrow 0$ for $|{z}|\rightarrow\infty$ is
asymptotically satisfied.

{{\it {Existence.}}}  As we already know, the operator $\hat{H_{x}}$
has one negative eigenvalue.  We show that the operator
$\hat{H_{y}}\hat{H_{x}}$ has a negative eigenvalue, too. It is
possible to rewrite the equation $\hat{H_{y}}\hat{H_{x}}\psi= {\tilde
  {\lambda}}\psi$ in the form
\begin{equation}
{\hat{H}_{y}}^{\frac{1}{2}}\hat{H_{x}}
{\hat{H}_{y}}^{\frac{1}{2}}({\hat{H}_{y}}^{-\frac{1}{2}}\psi)=
{\tilde{\lambda}}({\hat{H}_{y}}^{-\frac{1}{2}}\psi)
\end{equation}
(the operator ${\hat{H_{y}}}$ has only positive eigenvalues, hence,
operators like $\hat{H_{y}}^{+\frac{1}{2}}$ or
${\hat{H_{y}}}^{-\frac{1}{2}}$ are well defined and Hermitian). The
operator ${\hat{H}_{y}}^{\frac{1}{2}}
\hat{H_{x}}{\hat{H}_{y}}^{\frac{1}{2}}$ is also Hermitian and its
lowest eigenvalue can be obtained by minimization of the functional
\begin{equation}
F[{\chi} ]=
\frac {\langle {\chi}| {\hat{H}_{y}}^{\frac{1}{2}}\hat{H_{x}}
{\hat{H}_{y}}^{\frac{1}{2}}|
{\chi} \rangle  }{\langle {\chi} |{\chi} \rangle  }
\end{equation}
in the space of functions with integrable modulus squared.  Using the
fact that ${H_{y}}^{\frac{1}{2}}$ is self-conjugate and defining
${\phi} ={\hat{H_{y}}}^{\frac{1}{2}}{\chi} ,$ the functional can be
rewritten as
\begin{equation}
F[{\phi} ]=
\frac{\langle { \phi}|\hat{H_{x}}|{\phi} \rangle  }
{\langle {\hat{H}_{y}}^{-\frac{1}{2}}
{\phi}|{\hat{H}_{y}}^{-\frac{1}{2}}{\phi}
\rangle  }=\frac{\langle {\phi} |\hat{H_{x}}| {\phi} \rangle  }
{\langle {\phi} | {\hat{H_{y}}}^{-1}|{\phi} \rangle  }.
\label{32}
\end{equation}
The operator $\hat{H_{x}}$ has a negative eigenvalue, i.~e., there is
a function $| {\phi} \rangle$ such that $\langle {\phi} |\hat{H_{x}}|
{\phi} \rangle <0.$ The operator $\hat{H}_{y}$ has only positive
eigenvalues, i.~e., the form $\langle {\phi} |
{\hat{H_{y}}}^{-1}|{\phi} \rangle $ is always positive. Consequently
we constructed the Rayleigh-Ritz principle for the equation
$\hat{H_{y}}\hat{H_{x}}{\psi} ={\tilde{\lambda}}{\psi} $ and proved
that there is a function with a negative average value.  The
eigenvalue ${\tilde \lambda}_{-1}$ is even lower, thus we have
demonstrated the existence of a negative eigenvalue for the problem
(\ref{Hall}).

{\it Uniqueness.} Let us rewrite the operator $\hat{H_{x}}$ in the
following form
\begin{equation}
\hat{H_{x}}=\sum_{\alpha}{\lambda}_{\alpha} |\alpha \rangle
\langle \alpha |=
\sum_{\alpha\not= -1}{ \lambda}_{\alpha} |\alpha \rangle  \langle \alpha |
+{\lambda} _{-1} |-1\rangle  \langle -1|,
\end{equation}
where $\lambda_{\alpha}$ are the eigenvalues of ${\hat H}_{x},$
${\lambda}_ {-1}$ is the lowest eigenvalue of the operator
$\hat{H_{x}},$ the vector $|-1\rangle$ denoting the ``ground state''of
the operator ${\hat H}_{x}.$ The operator ${\hat{H}_{y}}^{\frac{1}{2}}
\hat{H_{x}}{\hat{H}_{y}}^{\frac{1}{2}}$ can be rewritten in the form
\begin{equation}
{\hat{H}_{y}}^{\frac{1}{2}}
\hat{H_{x}}{\hat{H}_{y}}^{\frac{1}{2}}=
{\hat{H}_{y}}^{\frac{1}{2}}\sum_{\alpha\not= -1}
{\lambda}_ {\alpha} |\alpha \rangle  \langle \alpha |{\hat{H}_{y}}
^{\frac{1}{2}}+{\hat{H}_{y}}^{\frac{1}{2}}
{\lambda}_{-1} |-1\rangle  \langle -1|{\hat{H}_{y}}^{\frac{1}{2}}\equiv
{\hat A}+{\hat B}.
\end{equation}
The image of the operator ${\hat B}={\hat{H}_{y}}^{\frac{1}{2}}
{\lambda}_ {-1} |-1\rangle \langle -1| {\hat{H}_{y}}^{\frac{1}{2}}$ is
one-dimensional, i.~e., ${\hat B}$ is a self-adjoint projector of rank
1. The operator
\begin{equation}
{\hat A}=
{\hat{H}_{y}}^{\frac{1}{2}}
\left (
\sum_{\alpha\not= -1}{\lambda}_{\alpha} |\alpha
 \rangle  \langle \alpha | + 0\cdot|-1\rangle  \langle -1|  \right )
{\hat{H}_{y}}^{\frac{1}{2}}
\end{equation}
is a nonnegative Hermitian operator with two zero eigenvalues.  We
perturb the operator ${\hat A}$ with a projector of rank 1 and make
use the following theorem\cite{Birman}

{\it {\bf Theorem.}  Let ${\hat A}$ be a self-adjoint operator and
  ${\hat B}$ a self-adjoint operator of rank 1. Take an interval $D$
  on the real axis and denote by $n(D)$ the number of eigenvalues of
  ${\hat A}$ in this interval. Then the number $m(D)$ of eigenvalues
  of ${\hat A}+{\hat B}$ within $D$ satisfies the estimate: }
\begin{equation}
n(D)-1 \leq m(D) \leq n(D)+1.
\end{equation}
Applying this theorem to the interval $(-\infty ,0)$ (we exclude
zero!) one can easily see that the operator $
{\hat{H}_{y}}^{\frac{1}{2}}\hat{H_{x}}{\hat{H}_{y}}^ {\frac{1}{2}}$
has not more than one negative eigenvalue.  But the eigenvalues of
this operator are the same as those of the operator
${\hat{H}_{y}}\hat{H_{x}}$ so that our problem has only one negative
eigenvalue.

\section{Negative eigenvalue problem}
\label{loww}
We construct upper and lower bounds for the negative eigenvalue
${\tilde \lambda}_{-1}$ of the operator ${\hat H}_{x}{\hat H}_{y}.$ In
the limit $F\rightarrow 0$ the lowest eigenvalues of the operators
${\hat H}_{x}$ and ${\hat H}_{y}$ do not depend on the detailed form
of the pinning potential\cite{Kramer1}.  This is due to the rapid
decay of the eigenfunctions in the region $|z|>D,$ where the form of
the potential is irrelevant. The lowest eigenvalue of the operator
${\hat H}_{x}$ is equal to $\lambda_{-1}=-{\mu^{2}F^{2}}/{2V_{0}},$
with $\mu$ the root of $\mu\tanh{\mu}=1$ (see Refs.\cite{Kramer1}).
In the region $|z|<D$ the normalized ``ground-state'' wave-function of
the operator ${\hat H}_{x}$ is
\begin{equation}
\phi_{-1}(z)=\frac{\cosh{\left (\sqrt{\frac{|{\lambda}_{-1}|}{\epsilon}}
z\right )}}
{\sqrt{D}\cosh{\left (\sqrt{\frac{|{\lambda}_{-1}|}{\epsilon}}D\right )}}.
\end{equation}
As the characteristic depth $\kappa$ of the ``potential well''
$U_{y}(z)$ is much larger than the ``size quantization energy''
${\epsilon}/{D^{2}},$ see Fig.~2, we can approximate the
eigenvalues and eigenfunctions of the operator ${\hat H}_{y}$ by those
of the infinitely deep quantum well
\begin{equation}
\lambda_{n}^{\prime}=\frac{\pi^{2}F^{2}}{8V_{0}}n^{2},\quad
\phi_{n}^{\prime}(z)=\frac{1}{\sqrt{D}}
\cos{\left (\frac{\pi n}{2D}z+\frac{\pi}{2}(n-1)\right )},\quad n=1,2\dots .
\label{evef}
\end{equation}
Using the variational principle (\ref{32}) we can write
\begin{equation}
{\tilde \lambda}_{-1}\le
\frac
{\langle \phi_{-1}|{\hat H}_{x}|\phi_{-1}\rangle}
{\langle \phi_{-1}|{\hat H}_{y}^{-1}|\phi_{-1}\rangle}=
-\frac{\frac{\mu^{2}F^{2}}{2V_{0}}}
{\sum\limits_{m}
\frac{{|{\langle \phi_{-1}| \phi_{m}^{\prime}\rangle}|}^{2}}
{\lambda^{\prime}_{m}}}.
\label{varnew}
\end{equation}
Substituting the eigenvalues and eigenfunctions (\ref{evef}) into
Eq.~(\ref{varnew}) we obtain
\begin{equation}
{\tilde \lambda}_{-1}\le
-\frac
{\frac{\mu^{2}F^{4}}{2V_{0}^{2}}}
{\sum\limits_{m=0}^{\infty}
\frac{8}{{\left (\mu^{2}+{\pi^{2}{(2m+1)}^{2}}/{4}\right )}^{2}}}
=-1.3292\frac{F^{4}}{V_{0}^{2}}.
\end{equation}

In order to obtain a lower estimate ${\tilde \lambda_{-1}}\ge B
>-\infty,$ (see (\ref{32})) we use the following inequality
\begin{equation}
{\tilde \lambda}_{-1}={\rm min_{\phi\in{\cal L}^{2}}} F[\phi ]\ge
{\rm min_{\Psi\in{\cal D}}}
\frac{\lambda_{-1}}{\langle \phi|{\hat H}_{y}^{-1}|\phi\rangle },
\label{supernew}
\end{equation}
where the functions belonging to ${\cal D}$ satisfy the conditions
{\it i}) $\langle \phi |\phi\rangle =1,$ {\it ii}) $\langle \phi
|{\hat H}_{x}|\phi\rangle <0,$ and {\it iii}) $\phi (z)=\phi (-z).$
The last condition is a consequence of the parity symmetry of the
operator ${\hat H}_{x}{\hat H}_{y}.$ Following the discussion in
section \ref{un}, the operator ${\hat H}_{x}$ has one negative
eigenvalue with an even eigenfunction.  For any odd function $\phi,$
$\langle\phi |{\hat H}_{x}|\phi\rangle\ge 0,$ thus an odd function
cannot minimize the functional (\ref{32}), hence the eigenfunction
corresponding to the lowest eigenvalue of the operator ${\hat
  H}_{x}{\hat H}_{y}$ has to be even.  Our goal is to find some
(large) constant $b$ such that $\langle \Psi |{\hat
  H}_{y}^{-1}|\Psi\rangle >b>0$ for any $\Psi\in{\cal D}.$ We rewrite
\begin{equation}
\langle \phi |{\hat H}_{y}^{-1}|\phi\rangle =
\sum\limits_{m}
\frac
{{|\langle \phi |\phi_{m}^{\prime}\rangle |}^{2}}
{\lambda^{\prime}_{m}}\ge 
\frac{{|\langle \phi |\phi_{1}^{\prime}\rangle |}^{2}}
{\lambda^{\prime}_{1}}
\label{AAA6}
\end{equation}
and expanding $\phi$ in the eigenfunctions of the operator ${\hat
  H}_{x},$ $\phi=\sum_{n=-1}c_{n}\phi_{n}\equiv c_{-1}\phi_{-1}
+{\tilde \phi},$ we can write
\begin{eqnarray}
{{|\langle \phi |\phi_{1}^{\prime}\rangle |}^{2}}
& = &
c_{-1}^{2}{{|\langle \phi_{-1}^{\prime} 
|\phi_{1}^{\prime}\rangle|}^{2}}
+
{{|\langle {\tilde \phi} |\phi_{1}^{\prime}\rangle |}^{2}}
+
2c_{-1}\langle \phi_{-1} |\phi_{1}^{\prime}\rangle 
\langle {\tilde \phi} |\phi_{1}^{\prime}\rangle 
\nonumber\\
& \ge &
c_{-1}^{2}{{|\langle \phi_{-1} |\phi_{1}^{\prime}\rangle |}^{2}}+
{{|\langle {\tilde \phi} |\phi_{1}^{\prime}\rangle |}^{2}}-
2|c_{-1}||\langle \phi_{-1} |\phi_{1}^{\prime}\rangle | 
|\langle {\tilde \phi} |\phi_{1}^{\prime}\rangle |. 
\end{eqnarray}
As $\langle \phi |{\hat H}_{x}|\phi \rangle <0,$
$-c_{-1}^{2}|\lambda_{-1}|+c_{1}^{2}\lambda_{1}+
c_{2}^{2}\lambda_{2}+\dots <0,$ and taking into account that
$\sum_{n}c_{n}^{2} =1$ we obtain
\begin{equation}
1>|c_{-1}|>\sqrt{\frac{\lambda_{1}}{|\lambda_{-1}|+\lambda_{1}}}
=0.9191,
\end{equation}
where we used the fact that $\lambda_{-1}=-1.4392{\epsilon}/{D^{2}}$
and $\lambda_{1}=7.8309{\epsilon}/{D^{2}}$ (see
Refs.\cite{Skvortsov,Kramer1}).  As $|\langle {\tilde \phi}
|\phi_{1}^{\prime}\rangle |\le ||{\tilde \phi }||\thinspace
||\phi_{1}^{\prime}|| \le\sqrt{1-c_{-1}^{2}} =0.3941,$ we obtain
${|\langle { \phi} |\phi_{1}^{\prime}\rangle |}^{2}\ge f(|c_{-1}|,K),$
where
\begin{equation}
f(|c_{-1}|,K)=
c_{-1}^{2}{{|\langle \phi_{-1} |\phi_{1}^{\prime}\rangle |}^{2}}+
K^{2}-
2|c_{-1}||\langle \phi_{-1} |\phi_{1}^{\prime}\rangle | K.
\end{equation}
We note that $|c_{-1}|\in [0.9191,1]$ and $K\in [0,0.3941].$ A simple
analysis shows that the minimal value of $f(|c_{-1}|,K)$ is equal to
$f(0.9191,0.3941)=0.119$ and substituting this value into
Eqs.~(\ref{supernew}) and (\ref{AAA6}) we obtain
\begin{equation}
{\tilde \lambda}_{-1}\ge -7.4603\frac{F^{4}}{V_{0}^{2}}.
\end{equation}
Hence, we have shown that in the limit $F\rightarrow 0$ the lowest
eigenvalue ${\tilde\lambda}_{-1}$ of the operator ${\hat H}_{x}{\hat
  H}_{y},$ ${\tilde \lambda}_{-1}$ behaves as $-{\gamma
  F^{4}}/{V_{0}^{2}}$ with $\gamma\in[1.3292,7.4603].$ The temperature
$T_{0}$ and the eigenvalue ${\tilde \lambda}_{-1}$ are connected by
the equation $T_{0}={\hbar\sqrt{|{\tilde \lambda}_{-1}|}}/
{2\pi\alpha}$ and we obtain the estimate
\begin{equation}
T_{0}=\frac{\xi}{2\pi}
\frac{\hbar F^{2}}{\alpha V_{0}}
\end{equation}
with $\xi\in [1.1530,2.7314].$

\end{appendix}



\end{document}